\documentclass[letterpaper,twocolumn,aps]{revtex4}

\usepackage{graphicx}

\begin{document}

\title{ARPES on Na$_{0.6}$CoO$_{2}$: Fermi surface, extended flat dispersion, and unusual band splitting}

\author{H.-B. Yang,$^{1}$ S.-C. Wang,$^{1}$ A.K.P. Sekharan,$^{1}$ 
H. Matsui,$^{2}$ S. Souma,$^{2}$ T. Sato,$^{2}$ T. Takahashi,$^{2}$ 
T. Takeuchi,$^{3}$ J.C. Campuzano,$^{4}$ R. Jin,$^{5}$ B.C. Sales,$^{5}$ D. Mandrus,$^{5}$
 Z. Wang,$^{1}$ H. Ding$^{1}$}

\affiliation{
(1) Department of Physics, Boston College, Chestnut Hill, MA 02467\\
(2) Department of Physics, Tohoku University, 980-8578 Sendi, Japan \\
(3) Research center for advanced waste and emission management, Nagoya 
University, Japan\\
(4) Department of Physics, University of Illinois at Chicago, Chicago, IL 60607 \\
(5) Condensed Matter Science Division, Oak Ridge National Laboratory, Oak Ridge, TN 37831
}
\begin{abstract}
\noindent The electronic structure of single crystals Na$_{0.6}$CoO$_2$, which are closely related to the superconducting Na$_{0.3}$CoO$_2$.$y$H$_2$O ($T_c \sim 5K$), is studied by angle-resolved photoelectron spectroscopy. While the measured Fermi surface is found to be consistent with the prediction of a local density band theory, the energy dispersion is highly renormalized, with an anisotropy along the two principle axes ($\Gamma$-$K$, $\Gamma$-$M$).  Our ARPES result also indicates that an extended flat band is formed slightly above $E_F$ along $\Gamma$-$K$.  In addition, an unusual band splitting is observed in the vicinity of the Fermi surface along the $\Gamma$-$M$ direction, which differs from the predicted bilayer splitting.

\noindent 


\end{abstract}
\maketitle

The recent discovery of superconductivity in Na$_x$CoO$_2$.$y$H$_2$O ($T_{c} \sim$  5 $K$) \cite{Discovery} has generated great interests in the condensed matter physics community due to its potential connection to the high-$T_{c}$ superconductivity. Over the past 17 years, the study of high-$T_{c}$ superconductivity has focused on copper oxide compounds (cuprates). The unexpected finding of superconductivity in cobalt oxide (cobaltates) has raised the hope that it may help solve the high-$T_{c}$ problem. Similar to the cuprates, Na$_x$CoO$_2$ has a layered structure. However, unlike the cuprates that have a square lattice in the plane, the cobaltate has a hexagonal lattice. For the electronic structure,  both compounds have partially filled 3$d$ orbitals, which form the low-energy excitations. In cuprates, Cu$^{2+}$ has 3$d^9$ configuration, and forms the highest $e_g$ ($d_{x^2-y^2}$) band near $E_F$, with a strong hybridization to O 2$p$ orbital. In cobaltates, the electronic configuration of Co$^{4+}$ is 3$d^5$, which occupies three lower $t_{2g}$ bands, with the topmost band being $A_{1g} = (d_{xy}+d_{yz}+d_{zx})/\sqrt{3}$. The hybridization to the O 2$p$ is greatly reduced due to the weaker overlap of the triangular bonding and the fact that the relevant orbitals in cobaltates are $t_{2g}$ rather than $e_g$ as in the cuprates.

In addition to these similarities in crystal structure and band orbitals, another important connection between the two materials is that both of them have strong electron correlations.  It is widely believed that the physics of the high-$T_{c}$ cuprates is that of a doped Mott insulator. It is believed that the cobaltate is also an electron-doped Mott insulator.  The spin configuration in the frustrated half-filled triangular lattice may favor a RVB state \cite{Baskaran}, and superconductivity in the vicinity of such the RVB state may prefer a $d$-wave order parameter \cite{Wang}, similar to the cuprates.   Moreover, the superconducting phase diagram of  Na$_x$CoO$_2$.1.3H$_2$O is found to have a dome-shape \cite{phasediagram}, which is also similar to the cuprates.

There also exist similarities between Na$_x$CoO$_2$ and another layered oxide superconductor Sr$_2$RuO$_4$ (ruthenate), which is believed to be a $p$-wave superconductor.  The two materials have similar behaviors in transport and magnetic susceptibility \cite{Terasaki,Maeno}.  In addition, both materials posses ferromagnetic fluctuations at low temperatures \cite{Imai,Ishida},  which are believed to be responsible for the $p$-wave pairing in the ruthenate. Thus a $p$-wave triplet superconducting state has been suggested for the cobaltate \cite{Singh2}. In addition, we note that both materials have the same electron filling level ($\frac{4}{3}$) for the maximum superconductivity. This is further away from half-filling than cuprates, in which the filling number of 1 $\pm$ 0.16 is usually regarded as the optimal doping level.

Band structure and Fermi surface (FS) topology are known to be important  for understanding unconventional superconductivity.  Different electronic structures may induce different fluctuations, which can lead to different pairing symmetry. In this Letter, we report an angle-resolved photoelectron spectroscopy (ARPES) study on Na$_{0.6}$CoO$_2$ single crystals. Although Na$_{0.6}$CoO$_2$ is not a superconductor, a reduction of Na concentration and proper hydration can turn this material into a superconductor \cite{Discovery}. It is worth mentioning that Na$_{0.5}$CoO$_2$ has a large thermoelectric power ($S \sim$ 100 $mV/K$ at 300 $K$), which is one order of magnitude larger than a typical $S$ of metals and high-$T_c$ cuprates \cite{Terasaki}. In our ARPES study, we observe clear band dispersion and FS crossings in Na$_{0.6}$CoO$_2$.  While the observed FS location is in a good  agreement with the one predicted by LDA band theory \cite{Singh_FS}, the observed bandwidth for the near-$E_F$ band is renormalized and anisotropic along the two principle axes ($\Gamma$-$M$, $\Gamma$-$K$).  The low-energy dispersion along $\Gamma$-$M$ can be described by a simple tight binding fit with a hopping integral $t \sim$ 44 $meV$.  The dispersion along $\Gamma$-$K$ is more complicated, with an extended flat band slightly above $E_F$ and a break in dispersion around 70 $meV$. A tight binding fit yields $t \approx$ 12 - 26 $meV$, indicating a mass renormalization of a factor of 5 - 10. In addition, we observe an unusual band splitting along $\Gamma$-$M$ , which is however absent along $\Gamma$-$K$.

High-quality Na$_x$CoO$_2$ single crystals are grown using
the floating-zone method. The Na composition $x \approx$ 0.6, determined by inductively coupled plasma atomic emission spectroscopy. ARPES experiments
are performed at the Synchrotron Radiation Center, Wisconsin. 
High-resolution undulator beamlines and Scienta analyzers with a capability of  
multi-angle detection have been used. Most spectra are measured using 22 $eV$ photons. 
The energy resolution is $\sim$ 10 - 20 $meV$, and the momentum resolution
$\sim0.02$ \AA$^{-1}$. Samples are cleaved and measured 
\emph{in situ} in a vacuum better than $8\times10^{-11}$ $Torr$ at low temperatures (20 - 40 $K$) with a flat (001) surface. The sample is stable and shows no sign of degradation during a typical measurement period of 12 hours. The sample is oriented according to its Laue diffraction pattern, which shows a clear hexagonal lattice.  LEED images obtained on fresh surface of cleaved samples also display a good hexagonal symmetry \cite{Zhang}.

\begin{figure}[{here}]
\includegraphics[ width = 8cm]{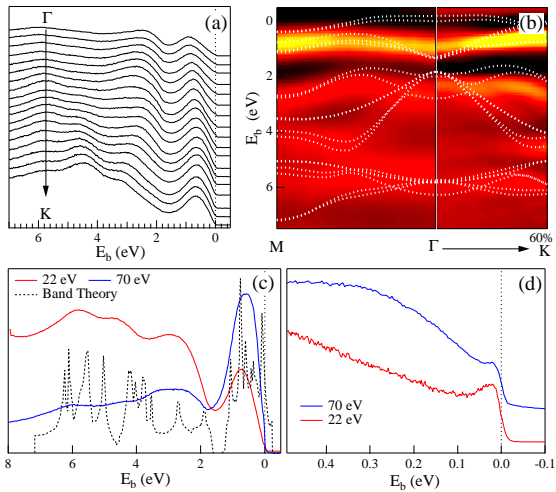}
\vspace{-10pt}
\caption{
Valence band of Na$_{0.6}$CoO$_{2}$. (a) EDCs along $\Gamma$-$K$ ($h\nu$ = 22 $eV$). (b) Intensity plots of the second derivatives of spectra along $\Gamma$-$M$ and $\Gamma$-$K$.  (c) Integrated spectra over a large $k$-space for 22 $eV$ (red) and 70 $eV$ (blue) photons. The black dashed line is the total DOS from band theory \cite{Singh_FS}. (d) Near-$E_F$ EDCs at $k_F$ along $\Gamma$-$K$ using 22 $eV$ (red) and 70 $eV$ (blue) photons. 
}
\label{VB}
\vspace{-5pt}
\end{figure}

In ARPES, a sign of a good surface is a clear valence band dispersion, which is observed in this material, as shown in Fig.~1. We measure the valence band at a number of photon energies, ranging from 16 - 110 $eV$.  We find that, at low photon energies, the band dispersion is clearly visible, as shown in Fig.~1a where a set of energy distribution curves (EDCs) along $\Gamma - K$ are measured with  22-$eV$ photons. In Fig.~1b, we plot the intensity of the second derivatives of the measured EDCs to display band dispersion.  It can be seen in Fig.~1b that the extracted band dispersion has many similarities with the calculated band structure \cite{Singh_FS}. However, the dispersion of the valence bands becomes less observable at relatively high photon energies ( $h\nu >$ 70 $eV$). In Fig.~1c, we integrate spectra over a large $k$-space to mimic the density of states (DOS) for two different photon energies (22 and 70 $eV$). Four peaks can be identified at binding energies of 5.9, 4.6, 2.8, 0.7 $eV$ respectively, matching well with the band calculation shown in Fig.~1c \cite{Singh_FS}. The band calculation shows that O 2$p$ bands (2 to 7 $eV$) and Co 3$d$ bands (below 2 $eV$) are well separated due to a weak Co $d$ - O $p$ hybridization. This is clearly reflected in Fig.~1c. In addition, a large enhancement of intensity for the 0.7-$eV$ peak with 70-$eV$ photons supports the Co 3$d$ characters for this peak, since 70 $eV$ is just slightly above the Co 3$p$ - 3$d$ resonant excitation ($\sim$ 63 $eV$). 

By examining the spectra in the vicinity of $E_F$, as shown in Fig.~1d, one can observe a weak but discernible peak sitting on a large ``background'' which is mainly a tail of the 0.7-$eV$ peak, similar to an earlier ARPES measurement in a similar material \cite{Valla}. According to the band calculation, this peak is the Co $A_{1g}$ band  which is at the top of Co $t_{2g}$ manifold and forms a Fermi surface, and the large background is mostly $t_{2g}$ bands.  We find, as shown in Fig.~1d, that  the Co 3$p$ - 3$d$ resonant excitation does not help the enhancement of the $A_{1g}$ peak since the resonance also enhances the background which has the same Co 3$d$ character.  Experimentally, we find that 22-$eV$ photons yield the best result in identifying this near-$E_F$ band.

\begin{figure}[{top}]
\includegraphics[width = 8cm]{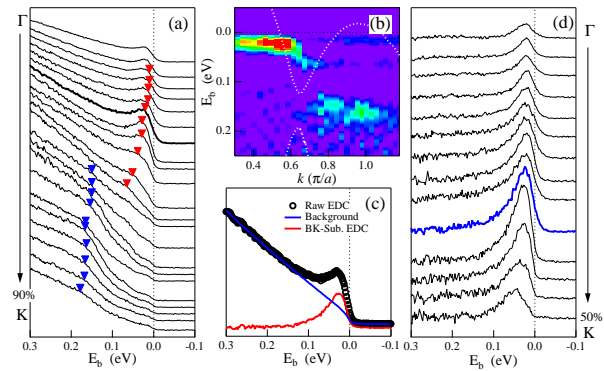}
\vspace{-10pt}
\caption{
Near-$E_F$ spectra at 40 $K$ along $\Gamma$-$K$ ($h\nu$ = 22 $eV$).  (a) EDCs along $\Gamma$-$K$. (b) Intensity plot of the second derivatives of the spectra shown in Fig.~2a. The white dashed lines are from the band calculation \cite{Singh_FS}. (c) An example of background subtraction. (d) EDCs after background subtraction. The blue curve is at $k_F$ = (0, 0.5).
}
\label{Crossing}
\vspace{-15pt}
\end{figure}

Figure 2 shows dispersing spectra along a principle axis $\Gamma$-$K$, or (0,0) - (0,$\frac{4}{3}$), in the hexagonal reciprocal lattice. The unit of $k$, used throughout this article, is $\pi/a$, where $a$ (2.84 \AA) is the nearest Co-Co distance.  As seen in Fig.~2a, a ``shoulder-like'' peak centered around $\sim$ 65 $meV$ starts to emerge at $k \sim$ (0, 0.66). With decreasing $k_y$, this peak disperses toward $E_F$, 
and seems to cross $E_F$ at $\sim$ (0, 0.5). Note that another broad peak around 160 - 200 $meV$ emerges at $k_y > 0.8$, which has no smooth connection to the $A_{1g}$ band. A careful examination of the spectra indicates that there is a ``break'' in dispersion, as shown in Fig.~2b. Comparing to the calculated band dispersion along $\Gamma$-$K$, this ``break'' occurs at almost the same $k$-location where the $A_{1g}$ band intersects with another $t_{2g}$ band. Therefore, this ``break'' might be a consequence of the mixing of the two bands. The second $t_{2g}$ band is predicted to cross the Fermi energy and form a small Fermi surface pocket near the $K$ point. However, no evidence of this FS crossing is observed in our experiment, which could easily result from a slight shift in the chemical potential. Another possible explanation for the ``break'' in dispersion is that the electron band interacts with collective modes, such as phonon modes. Three phonon modes, with the energy of $\sim$ 55 - 75 $meV$, have been observed by a Raman experiment \cite{Iliev}. 

To see the low-energy excitation more clearly, we subtract the large ``background'' from an EDC.  One example of the subtraction is demonstrated in Fig.~2c where an almost straight line is used for the background. The background-subtracted EDCs along $\Gamma$-$K$ are plotted in Fig.~2d. One can clearly see that an asymmetrical peak disperses toward $E_F$ while its lineshape sharpens up. At $k \approx$ (0, 0.5),  the peak experiences a substantial reduction in its intensity, indicating that the peak is crossing $E_F$ and the sharp Fermi function removes most of spectral weight above the $E_F$. After the crossing, a small peak near $E_F$ persists for a extended range of $k$, from $\sim$(0, 0.5) to (0,0). This indicates that there is an extended flat band just above the $E_{F}$. The small peak is the leftover intensity from the Fermi function cutoff. We note that there is an extended flat band in both cuprates (i.e. Bi$_2$Sr$_2$CaCu$_2$O$_8$) and ruthenate Sr$_2$RuO$_4$.  While the influence of the flat band to superconductivity is still under debate, the enhanced DOS near $E_F$ would certainly enhance spin and charge fluctuations, which are found to be strong in all of the three materials.

\begin{figure}[{top}]
\includegraphics[width = 8cm]{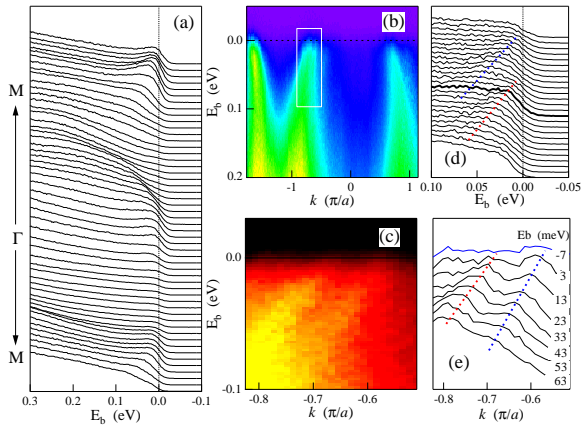}
\vspace{-10pt}
\caption{
Near-$E_F$ spectra at 40 $K$ along $\Gamma$-$M$.(a) EDCs and (b) $E$-$k$ intensity plot for a long cut along $M$-$\Gamma$-$M$.  Plots of (c) $E$-$k$ intensity, (d) EDCs, (e) MDCs magnified from the white box in Fig.~3b. Dashed lines are the guides for peak positions.
}
\label{FS}
\vspace{-15pt}
\end{figure}

ARPES spectra along another principle axis $\Gamma$-$M$, or (0,0) - ($\frac{2}{\sqrt{3}}$, 0), are plotted in Fig.~3. Symmetric dispersion can be observed along a long cut in $k$ over different  Brillouin zones (BZs). We observe some differences for spectra between $\Gamma$-$M$ and $\Gamma$-$K$. One major difference is that there is a band splitting along $\Gamma$-$M$. This splitting can be easily seen if we zoom in at the band crossing, as shown in Fig.~3 (c) to (e) where we plot an intensity plot, EDCs, and momentum distribution curves (MDCs) in the vicinity of the band splitting.  From these three displays, we observe two nearly parallel bands in the vicinity of $k_F$ with an energy separation of $\sim$ 60 $meV$ and a momentum separation of $\sim$ 0.1 $\pi/a$.
 
We summarize our ARPES results in Figure 4.  In Fig.~4a, we plot the FS crossings (FSCs) in the hexagonal BZ. For comparison, the calculated FS of Na$_{0.5}$CoO$_2$ is also plotted in Fig.~4a \cite{Singh_FS}. We have measured many samples on several photon energies,  with consistent results. All the FSCs shown in Fig.~4a are extracted directly from measurements, and no symmetry operations have been applied here. Note that the FSCs with same symbols over different BZs are obtained from a same sample during a short time interval. This eliminates potential problems from sample misalignment and surface contamination, and helps to accurately determine the size of the FS. As seen from Fig.~4a, the size of measured FS in Na$_{0.6}$CoO$_2$ is slightly larger than the calculated one in Na$_{0.5}$CoO$_2$.  We also notice that in a recent ARPES study for a more Na-doped sample (Na$_{0.7}$CoO$_2$), the observed hole-like FS appears larger than the one observed here \cite{Hasan}. The conclusion of having a smaller occupied area for more doped electrons from Na apparently violates the Luttinger theorem.  It is possible that some electron carriers become localized at higher Na doping levels, resulting a reduction in the number of itinerant electrons which contribute the area of the FS. This might also explain the unusual ferromagnetic transition observed in highly Na-doped samples \cite{Motohashi,Gavilano}.

\begin{figure}[{top}]
\includegraphics[width = 8cm]{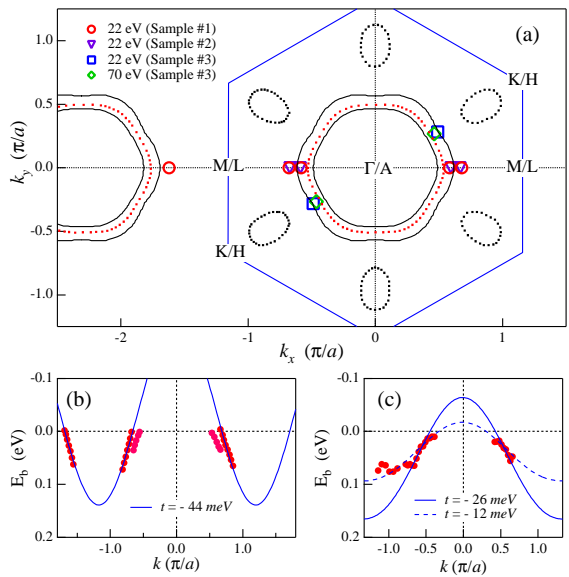}
\vspace{-10pt}
\caption{
(a) Measured FS crossings (symbols) comparing to the calculated FS in $k_z = 0$ (black solid lines) and $k_z = 0.5$ (red dashed lines) planes. The blue hexagon is the 2D Brillouin zone. (b)  Extracted band positions along $\Gamma$-$M$ (red dots) and a tight binding fit with $t$ = - 44 $meV$ (solid line)  (c) Extracted band positions along $\Gamma$-$K$ (red dots) and two tight binding fits with $t$ = - 12 $meV$ (solid line) and $t$ = - 26 $meV$ (dashed line). 
}
\label{FS}
\vspace{-15pt}
\end{figure}
 
We extract band dispersion from peak positions in MDCs. The extract band positions along $\Gamma$-$M$ and $\Gamma$-$K$ are plotted in Fig.~4b and c. Reliable values of the dispersion can be extracted from $E_F$ to $\sim$70 meV since beyond this energy the spectral linewidth becomes so broad that its position is ill-defined. As seen in Fig.~4b, the measured band dispersion along $\Gamma$-$M$ is nearly linear for the 70-$meV$ range, with a Fermi velocity $v_F \sim$ 0.41 $eV$\AA. We use a simple tight binding band for the triangular lattice to fit the measured band dispersion for the outer band,
\begin{equation}
\epsilon_k = -2t(\cos{k_x} + 2\cos{\frac{k_x}{2}}\cos{\frac{\sqrt{3}k_y}{2}}) - \mu
\end{equation}
where $t$ is the nearest neighbor hopping integrals, and $\mu$ the chemical potential. The fitting result yields $t \sim$ 44 $meV$ (with a negative sign) along $\Gamma-M$. The measured dispersion along $\Gamma$-$K$ is more complicated, as shown in Fig.~4c. In addition to the ``flat dispersion'' around 70 $meV$ due to the ``break'',  the dispersion changes its slope at $\sim$ 20 $meV$. This slope change may be a consequence of the flat band  observed in Fig.~2.
While a tight binding fit for the range between $E_F$ and 20 $meV$ yields $t \sim$ 12 $meV$ and $v_F \sim$ 0.12 $eV$\AA, another fit for the range between 20 to 70 $meV$ produces $t \sim$ 26 $meV$ and $v_F \sim$ 0.24 $eV$\AA. Both values of $t$ are smaller than the one along $\Gamma-M$. The different values of $t$ along the two principle axes are unexpected from a simple tight binding fit or LDA band theory. This indicates that either $t$ is $k$-dependent, or there may be other factors that modify the band dispersion, such as low-lying modes or energy gaps. 

The small value of $t$, especially along $\Gamma$-$K$ ($t \sim$ 12 - 26 $meV$),  is puzzling. It is estimated that the super exchange interaction $J$ is 12 and 24 $meV$ in Na$_{x}$CoO$_2$ \cite{Wang}. Apparently, the small $t$ observed here is the effective $t_{eff}$, which is renormalized from the original band $t$, estimated to be $\sim$ 130 $meV$ (the total band width is 12$t$ $\sim$ 1.6 $eV$) \cite{Singh_FS,Kunes}.  The reduction of $t$ indicates a strong mass renormalization of a factor of 5 - 10, which is likely due to the strong correlation in this material.  A similar mass renormalization ($\sim$ 7) is also suggested by a large electronic specific heat coefficient $\gamma$ ($\sim 48 mJ/molK^2$) observed in this material \cite{Ando,Chou}. We note that both the flat section of the FS and the extended flat band observed along $\Gamma$-$K$ would make it the dominant direction in contribution to the DOS.  Therefore, $t$ along $\Gamma$-$K$ is more relevant to the specific heat result.

Perhaps more surprising finding from this study is a relatively large band splitting along $\Gamma-M$.  While band theory predicts a bilayer splitting resulting from the two CoO$_2$ planes per unit cell \cite{Singh_FS}, we cast some doubts to this assignment due to the following reasons. (1) In the case of the bilayer splitting, the $c$-axis hopping integral $t_{\bot}$ would be 30 $meV$, similar to the in-plane $t$.  This is inconsistent with the large transport anisotropy ($\sigma_{ab} / \sigma_c \sim 200$ at 4 $K$) \cite{Terasaki}. (2) Perhaps more importantly, the two CoO$_2$ layers in a unit cell have identical geometry for Co atoms, and this would not generate a bilayer splitting for the Co $A_{1g}$ band near $E_F$ which has very small overlap with oxygen orbitals. In support of this, we note that the band calculation shows no discernable gap at the zone boundary of the doubled unit cell \cite{Singh_FS}. This suggest that the band can be unfolded to yield a $c$-axis dispersion rather than a splitting. (3)  No band splitting is observed along $\Gamma-K$, which is not consistent with a simple bilayer splitting, although different $k_z$ may modify the the size of splitting.  It is known that other phenomena, such as itinerant ferromagnetism,  can also cause band and FS splitting.  Several experiments have detected ferromagnetic fluctuations at low temperatures in this material \cite{Ishida, Jin}. Although speculative, it is possible that this fluctuation may order on the surface, causing a ferromagnetic band splitting.   More studies are necessary in order to resolve this issue.

We thank M. Fisher, C. Gundelach, and H. Hochst for technical support in synchrotron experiments, P.D. Johnson, P.A. Lee, N.P. Ong, D.J. Singh, and J.D. Zhang for useful discussions and suggestions.
This work is supported by NSF DMR-0072205, DOE DE-FG02-99ER45747, Petroleum Research Fund, Sloan Foundation. 
The Synchrotron Radiation Center is supported by NSF DMR-0084402. Oak Ridge National laboratory is managed by UT-Battelle, LLC, for the U.S. Department of Energy under contract
DE-AC05-00OR22725.


\vspace{-10pt}
\begin{thebibliography}{10}
\bibitem{Discovery} K. Takasa \textit{et al}., Nature \textbf{422}, 53 (2003).
\bibitem{Baskaran} G. Baskaran,  Cond-mat/0303649, 2003.
\bibitem{Wang} Q.H. Wang, D.H. Lee, and P.A. Lee, Cond-mat/0304377, 2003.
\bibitem{phasediagram} R.E. Schaak \textit{et al}.,  Nature \textbf{424}, 527 (2003).
\bibitem{Singh2} D. J. Singh, Phys. Rev. B \textbf{68}, 020503 (2003).
\bibitem{Terasaki} I. Terasaki, Y. Sasago, and K. Uchinokura, Phys. Rev. B \textbf{56}, R12685 (1997).
\bibitem{Maeno} Y. Maeno \textit{et al}., Nature \textbf{372}, 532 (1994).
\bibitem{Imai} T. Imai \textit{et al}., Phys. Rev. Lett. \textbf{81}, 3006 (1998).
\bibitem{Ishida} K. Ishida \textit{et al}., Cond-mat/0308506, 2003.
\bibitem{Zhang} J.D. Zhang, private communication.
\bibitem{Singh_FS} D. J. Singh,  Phys. Rev. B \textbf{61}, 13397 (2000).
\bibitem{Valla} T. Valla \textit{et al}., Nature \textbf{417}, 627 (2002).
\bibitem{Iliev} M.N. Iliev \textit{et al}., Cond-mat/0308065, 2003.
\bibitem{Kunes} J. Kune\u{s}, K.-W. Lee, and W.E. Pickett, Cond-mat/0308388, 2003.
\bibitem{Ando} Y. Ando \textit{et al}., Phys. Rev. B \textbf{60}, 10580 (1999).
\bibitem{Chou} F.C. Chou \textit{et al}., Cond-mat/0306659, 2003.
\bibitem{Ikeda} H. Ikeda, Y. Nisikawa, and K. Yamada, Cond-mat/0308472, 2003.
\bibitem{Jin} R. Jin \textit{et al}., Cond-mat/0306066, 2003, Phys. Rev. Lett., in press. 
\bibitem{Hasan} M.Z. Hasan \textit{et al}., Cond-mat/0308438, 2003.
\bibitem{Motohashi} T. Motohashi \textit{et al}., Phys. Rev. B \textbf{67}, 064406 (2003).
\bibitem{Gavilano} J.L. Gavilano \textit{et al}., Cond-mat/0308383, 2003.
\end{thebibliography}
\end{document}